# Climatological benchmarking of AI-generated tropical cyclones

**Yanmo Weng[1] and Avantika Gori[1]**

[1]Department of Civil and Environmental Engineering, Rice University, Houston, TX, USA.
Corresponding author: Yanmo Weng (yanmow@rice.edu)

**Key Points:**

- This study provides a comprehensive multi-year benchmarking of AI-generated tropical cyclones using historical reanalysis data.

- AI models simulate TC track, forward speed, and outer size well but underestimate storm intensity in some cases and overestimate inner size.

- Despite advances in TC intensity simulation in recent AI models, AI-generated TCs do not generally conform to gradient wind balance.

**Abstract**

This study presents a comprehensive climatological benchmarking of tropical cyclones (TCs) generated by AI-based global weather prediction models. Using all TC events from the North Atlantic and Western Pacific basins between 2020 and 2025, we assess the ability of two AI models (Pangu-Weather and Aurora) to reproduce observed TC track density, climatology of storm characteristics, and physical consistency with TC theory. By comparing AI-simulated TCs with ERA5 reanalysis, we benchmark the distributions of intensity, size, forward speed, and evaluate the model's ability to credibly simulate extratropical transition. Results show that both Pangu and Aurora perform well in reproducing storm track density, forward speed distribution, and outer size distribution. Aurora shows an improved performance in simulating storm intensity compared to Pangu, with less bias in the distribution of minimum central pressure and maximum wind speed. However, both models overestimate the distribution of storm inner size (radius of maximum winds), especially for extreme events. AI models capture the relative frequency and temporal evolution of extratropical transition patterns with reasonable accuracy. The AI-simulated TCs are also less likely to conform to gradient wind balance compared to ERA5, indicating that the AI TCs may not be physically realistic in many cases. This benchmarking identifies systematic biases that can guide future corrections and support extended applications of AI models for TC hazard and risk assessment. Our work establishes a foundation for future studies using AI weather models in the context of TC climatological and hazard research.

**Plain Language Summary**

Tropical cyclones can cause severe damage to coastal regions due to their wind, rain, and storm surge. New artificial intelligence (AI) weather models offer fast simulation of TCs, but the characteristics of AI TCs may be biased and may not satisfy physical constraints. This study implements a long-term evaluation of the Pangu-Weather and Aurora models by comparing their simulations initialized by historical TC locations to real storm data from 2020-2025. We find that Pangu-Weather tracks storm paths and sizes well but underestimates the strength of the most intense storms, while the Aurora model displays improved intensity simulation performance. We find that TCs generated from both Pangu-Weather and Aurora do not conform to well-known physical constraints, especially in regions close to the storm center. These findings help identify and correct model biases. With improved accuracy, AI-generated storms can support future research on rainfall, flooding, and disaster risks, offering a valuable tool for understanding and preparing for extreme weather events.

**1 Introduction**

Tropical cyclones (TCs), associated with high-speed winds, storm surges, and heavy rainfall, are extreme weather events that cause severe impacts to coastal areas across the US and around the world (Gori et al., 2022, 2025; Peduzzi et al., 2012). Accurate forecasting and simulation of TC tracks and storm characteristics are crucial for evaluating extreme wind, rainfall, and storm surge conditions in advance of landfalling storms and for reliable TC risk assessment. Numerical weather prediction (NWP) models are commonly used for weather forecasting (Bauer et al., 2015), but they are physically complex and computationally expensive. As a result, artificial intelligence (AI) and machine learning (ML) models have become popular alternatives for TC forecasting and simulation. State-of-the-art AI/ML applications can be categorized into four main areas (Chen et al., 2020): TC track forecasting (Racah et al., 2017; Song et al., 2022; Wu et al., 2022), TC intensity prediction (Bai et al., 2021; Boussioux et al., 2022; Varalakshmi et al., 2023), extreme rainfall prediction during TC events (Mikhaylov et al., 2024; Uddin et al., 2022), and incorporating ML to improve the NWP model performance (Jeong & Yi, 2023; Meng et al., 2024).

Typical ML models in the TC domain are often developed using task-specific architectures and trained on relatively small and domain-limited datasets. These models typically focus on a specific region, storm type, or storm variable (e.g., intensity or rainfall). As a result, they tend to perform well within the range of the training data but struggle to generalize to unseen or extreme cases, such as high-intensity storms or events with unusually large spatial extents (Frifra et al., 2024; Jeong & Yi, 2023; Jiang et al., 2018; Wu et al., 2022). In contrast, recent AI-based weather prediction models, especially those built using deep learning architectures such as transformers and graph neural networks, are designed for broader applicability (Bi et al., 2023; Lam et al., 2023; Pathak et al., 2022). These models are typically trained on massive, globally distributed datasets and are not restricted to TC-specific features or regions. This combination of advanced model architecture and diverse, long-term data enables them to learn complex spatiotemporal dependencies across a wide range of weather phenomena. Consequently, AI-based weather prediction models offer improved generalizability and scalability (Lam et al., 2023; Price et al., 2025). Although training such models requires significant computational resources, once trained, they can be fine-tuned efficiently and perform rapid inference for TC-related prediction as part of global weather forecasting.

Current AI-based models primarily focus on global weather forecasting, and their effectiveness in simulating a wide range of TC events remains unclear. Recent studies have started investigating the forecasting accuracy of AI-based weather models for TC events. For example, Xu et al. (2024) evaluated an AI-based model's performance for TC Khanun in 2023, demonstrating that the AI-generated predictions captured the TC center and key atmospheric systems more accurately than the Weather Research and Forecasting (Skamarock et al., 2008); WRF model. However, their study was limited to a single case study storm and did not address how to apply AI models to realistically simulate large event sets of TCs. Sun et al. (2024) conducted a comprehensive investigation of whether AI-based weather models can predict the intensity of extreme TC events (i.e., Category 3-5). They excluded the category 3-5 TC events, retrained the model, and then tested it on category 3-5 TC events. An important conclusion drawn from their study is that current AI-based weather models have poor predictions on out-of-distribution extreme TC events.

AI-based weather forecasting models can be applied for both operational forecasting and to generate large TC event sets for hazard/risk assessment. In operational forecasting, models can continuously assimilate new observational data to calibrate predictions back toward reality, effectively reducing error accumulation over time. In contrast, hazard and risk assessment applications require models to perform reliably in the absence of observational data constraints, generating realistic TC characteristics over extended simulation periods. The performance of AI-based weather models for TC risk assessment remains largely unexplored. This gap is particularly significant because TC hazard and risk assessment requires accurate representations of storm climatology across various intensity levels, spatial distributions, and temporal patterns. Without a comprehensive understanding of how these models perform in unconstrained simulations, their utility for synthetic track generation, hazard mapping, infrastructure design, and long-term risk management remains uncertain. One state-of-the-art study proposed by Jing et al. (2024) generates synthetic TCs using AI-based weather forecasting models and then downscales them to high-resolution. This approach could be used for follow-up hazard assessment. However, the bias of AI-based weather forecasting models is not corrected, and it is unclear if the synthetic TCs satisfy physical constraints and represent realistic storm climatology. Additionally, DeMaria et al., (2025) evaluate AI-based weather model performance on TC track and intensity forecast, finding that AI-based weather models show comparable or smaller track errors for TCs during the period May-November 2023 than numerical weather prediction models, while intensity forecast errors remain very large. However, due to the short TC period covered in the study, the intensity bias is not comprehensively quantified. While past studies (DeMaria et al., 2025; Liu et al., 2024) and climate emulators such as ACE2 (Watt-Meyer et al., 2025) have investigated the accuracy of AI-generated TCs for forecasting applications, there is a need to identify which atmospheric parameters are biased or unbiased in AI simulations, whether AI-generated TCs credibly capture dependence structures between different storm parameters, and whether they satisfy physical constraints. This paper addresses these gaps by providing a comprehensive comparison of TC climatology between reanalysis data and AI simulations, establishing a benchmark for future research utilizing AI-based weather forecasting model results and evaluating their potential application for broader TC hazard and risk assessment.

## 2 Data and Methods

### 2.1 AI model and TC simulation procedure

Commonly used global AI-based weather prediction models (AI NWPs) includes FourCastNet (Pathak et al., 2022), Graphcast (Lam et al., 2023), Pangu-Weather (Bi et al., 2023), and Aurora (Bodnar et al., 2025). Notably, Pangu-Weather (referred hereafter as Pangu) is the first AI NWP that has been shown to outperform dynamical weather models (e.g., WRF) for some cases (Bi et al., 2023; Liu et al., 2024; Shi et al., 2025; H. Xu et al., 2024), and has shown the highest accuracy in simulating TCs in past studies (Bi et al., 2023; Hua et al., 2025; Sahu et al., 2025). Besides Pangu, Aurora is the most recent AI NWP, which has not yet been systematically evaluated for its performance of TC simulation. Additionally, since the pre-trained Pangu and Aurora models are publicly available and the models are straightforward to implement for inference, we select both models to conduct the TC benchmarking and analysis. We use data from the European Center for Medium-Range Weather Forecasting's Reanalysis v5 (ERA5) as the primary benchmarking dataset since both AI models were trained extensively on ERA5 data. This benchmarking analysis can be applied to other AI weather models if the model inputs are ERA5 reanalysis data at 0.25° resolution and the model produces the same output variables as Pangu/Aurora (i.e., the variables shown in Figure 1). The Pangu model was pretrained with ERA5 data from 1979 to 2017 (Bi et al., 2023), while Aurora was pretrained with various datasets including ERA5 data from 1979 to 2019 (Bodnar et al., 2025).

To evaluate the model's ability to simulate TCs in the absence of continuously assimilated observational data, we conduct 7-day simulations for each historical storm occurring in the North Atlantic (NA) and Western North Pacific (WNP) ocean basins. We identify observed TC occurrences based on the International Best Track Archive for Climate Stewardship version 4, revision 0 (IBTrACS; Knapp et al., 2010). Figure 1 shows the flowchart of how the historical TCs are simulated and analyzed using each AI model. We select all TCs occurring from 2020 onward, which were not used for training either of the AI models (i.e., these are unseen events). Since Pangu was only trained with ERA5 data until 2017, we also include Pangu simulations for TCs occurring from 2018-2019, and these results are available in the Supporting Information (Figures

S13 – S16). For each TC, IBTrACS records storm position and characteristics at 3-hour intervals. We initialize each TC in the AI models when the storm first reaches Tropical Storm intensity (34 knots), as recorded in IBTrACS, and then simulate the track for seven days. We also conduct a sensitivity test, initializing each TC when it first reaches 25 knots, 50 knots, and 64 knots in IBTrACS.

Pangu and Aurora models use auto-regression inferencing to perform the multi-day simulations. For the Pangu model, both 6-hour and 24-hour pretrained weights are used within a single 7-day simulation. As shown in Figure 1, the atmosphere variables at the initial timestep $t_0$ are fed into the 6-hour pretrained model to make the prediction for $t_6$. Then $t_6$ serves as the input for predicting atmosphere variables at timestep $t_{12}$, again using the 6-hour model. At 24-hour interval (e.g., $t_{24}$, $t_{48}$, $t_{72}$), we switch to using the 24-hour pretrained model. For these specific timesteps, the model uses the previous 24-hour interval timestep as input and feeds it into the 24-hour model (e.g., $t_0 \rightarrow t_{24}$, $t_{24} \rightarrow t_{48}$). For all other timesteps, the 6-hour model is used. This hybrid inference approach reduces the accumulated error during the auto-regression process by leveraging the 24-hour model's ability to make direct longer-term predictions at key intervals (Bi et al., 2023; Couairon et al., 2024; Zhong et al., 2024). For the Aurora model, the atmosphere variables at the current timestep $t_0$ and previous 6-hour timestep ($t_{-6}$) are used as the input for the model. Aurora uses a single pre-trained model to simulate the weather for the full 168-hour (7-day) simulation in 6-hour increments. In our current TC simulations, ERA5 data is only used for the initial timestep. For the rest of the simulation, the Pangu and Aurora models predict subsequent timesteps based only on their own predictions from the previous time step. We compare the Pangu and Aurora simulation results with ERA5 in the post-processing step to evaluate similarities and differences in the simulated TC climatology over 5 years (2020-2025).

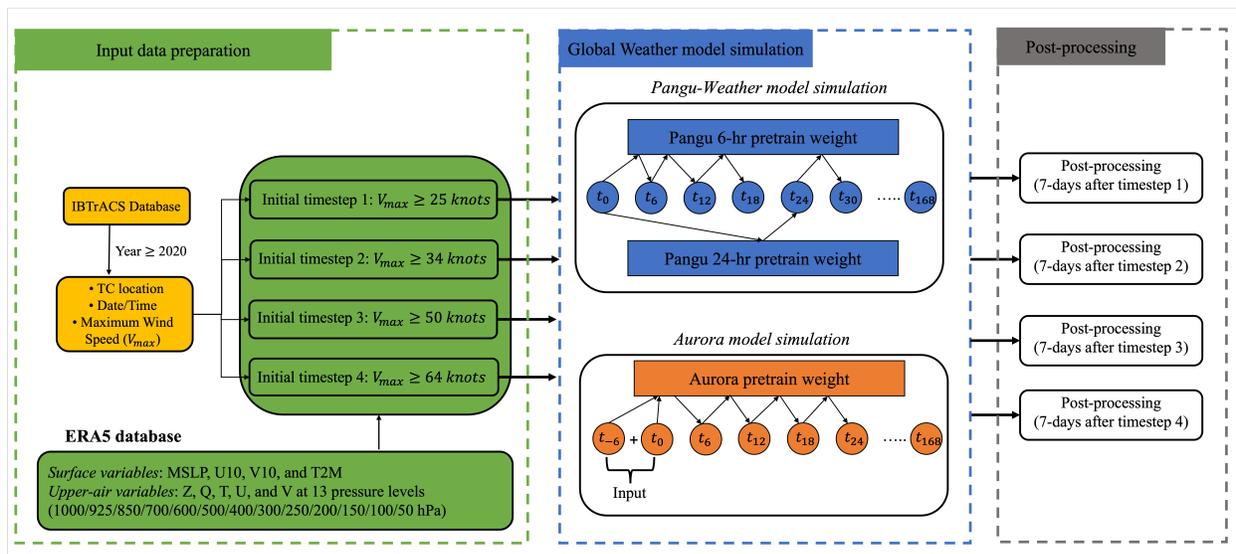

**Figure 1.** Pangu-Weather and Aurora model and post-processing implementation.

## 2.2 TC tracking within AI weather models

After obtaining the model predictions and corresponding ERA5 data for each TC simulation, the first step of extracting the storm characteristics in the post-processing procedure is to find the accurate TC center in both ERA5 and the AI simulation. For all ERA5 timesteps and for the first Pangu/Aurora timestep, the TC center in IBTrACS is used as the initial guess. Then the TC center is determined by searching for the minimum mean sea level pressure (MSLP) within a 445 km radius of this initial guess. This minimum MSLP location is assigned as the true TC center. For subsequent AI model timesteps, we use a combination of steering flow and past movement for the initial guess of the next track center (Bi et al., 2023; Grijn, 2002; Marchok, 2021; Zhong et al., 2024), since the simulated track may deviate substantially from observations and reanalysis. For the second AI model timestep, since there is no past movement information available, only the azimuthally averaged steering flow (calculated as the average of wind fields at 850, 700, 500, and 200 hPa within a 445 km radius of the previous true TC center) is used to estimate the next center location. From the third timestep onward, both the past movement (based on the previous two true TC centers) and the steering flow are weighted equally (0.5 each) to estimate the next TC center position. After estimating each position, the minimum MSLP within a 445 km radius of the estimated location is identified as the true TC center for that timestep. This process is repeated for the entire 7-day simulation period, allowing for accurate tracking of the TC position through time. This TC tracking method has been applied in multiple studies and compares well with other tracking methods (Chakraborty et al., 2020; Froude et al., 2007; Miyachi & Enomoto, 2021).   To ensure a valid TC system exists at each time step (i.e., the TC has not dissipated), three conditions must be satisfied starting from third timestep: 1) For the Northern Hemisphere, the 850 hPa relative vorticity must have a maximum value greater than $5 \times 10^{-5}$ within a radius of 278 km; 2) The minimum MSLP should be lower than 1010 hPa; and 3) When the cyclone is on land, the maximum 10-meter wind speed within a radius of 278 km from center should surpass 8 m/s. The TC track is terminated once any condition is not met (Bi et al., 2023). Notably, since the TC center at the first and second timestep is identified based on the minimum MSLP within a 445 km radius, a few initial cases may have minimum MSLP values greater than 1010 hPa. Additionally, to avoid very weak storms and ensure the TC track is long enough for extracting valuable information, cases with fewer than five valid data points in the entire 7-day simulation are also excluded. After applying these conditions, there are 90 and 97 storm events in the NA and WNP that satisfy the requirements.

TC tracks are visualized through track density maps; the NA and WNP basins are first divided into 2.5º latitude x 2.5º longitude cells and then we count the number of timesteps for which the TC centers of all storms are located in each cell. Besides visually inspecting the track density maps, we also conduct the False Discovery Rate (FDR) approach to identify if there are any geographic regions where the TC track tendencies are significantly different between AI simulations and ERA5. We conduct hypothesis testing in each grid cell and apply the FDR method to control the expected proportion of false discoveries among rejected hypotheses at the grid-cell level, and to determine the field significance of the track densities in each basin. The FDR method is robust to spatial correlation and provides better interpretability for climate/meteorology applications (Wilks, 2006). Fisher's exact test is performed in each cell to estimate the local p-value for the cell. Then, the Benjamini–Hochberg FDR (BH-FDR) rule is applied with the false discovery rate (q) set as 0.05 (Benjamini & Hochberg, 1995). The BH-FDR procedure ranks p-values from smallest to largest and rejects hypotheses up to the largest k, where $p(k) \leq (k/m)q$, and m is the total number of tests. The analysis returns the grid cells with significant differences between the AI model and ERA5 track densities.

## 2.3 Climatological Metrics

To evaluate whether the models accurately simulate the climatology of TCs, we select several storm characteristics to describe the TC. The minimum mean sea level pressure (MSLP) and maximum azimuthal-average azimuthal wind ($V_{max}$) are both used to describe the storm intensity distribution. The radius to the maximum azimuthal wind speed ($R_{max}$) and TC outer size (defined as the radius to a given weak azimuthal wind speed) are used to describe the TC size climatology. $R_{max}$, or inner size, is an important storm parameter that modulates the severity and extent of TC hazards, such as rainfall (Lavender & Mcbride, 2020) and storm surge (Irish et al., 2008). The outer size is defined as the radius at which the 10-m azimuthally averaged wind speed first falls below a certain low value (Chavas et al., 2016). Gori et al. (2023) estimated six outer size metrics, defined as the radii to azimuthal wind speed equal to 2, 4, 6, 8, 10, and 12 m/s ($r_2$ - $r_{12}$), in the ERA5 from 1950 to 2020. They found that the $r_6$ and $r_8$ size metrics in ERA5 yield good agreement with the measured TC size from satellite data. Therefore, we choose to utilize the $r_6$ and $r_8$ metrics to describe the TC outer size. We analyze these TC characteristics using univariate and bivariate probability density functions to visualize the distributional differences between AI models and ERA5. To statistically quantify whether observed distributional differences are significant, we apply the two-sample Kolmogorov-Smirnov (KS) test to evaluate whether two samples are drawn from the same underlying distribution by measuring the maximum distance between their cumulative distribution functions.

In this study, ERA5 is used as the primary reference for benchmarking because the AI weather models evaluated here are trained and optimized specifically against ERA5 reanalysis fields. Their architectures are designed to minimize discrepancies (e.g., mean squared error) relative to ERA5, making it the effective ground truth for model evaluation. While IBTrACS is considered the best estimate of historical TC parameters, using storm characteristic distributions from IBTrACS for comparison with the AI models could introduce inconsistencies since the discrepancies would encompass both the biases between ERA5 and IBTrACS and the biases between the AI models and ERA5. The systematic differences between ERA5 and IBTrACS, especially ERA5's underprediction of TC intensity and overprediction of TC inner size ($R_{max}$), have been discussed in previous studies (Bian et al., 2021; Gori et al., 2023). Here, we utilize TC parameter distributions from IBTrACS to provide additional context about the reliability of AI TC simulations, but we note that evaluating the AI models' performance relative to IBTrACS for characteristics such as MSLP, $V_{max}$ or $R_{max}$ would not provide a fair assessment of model performance, as the models are not trained toward that reference.

## 2.4 Gradient Wind Balance

Most AI weather models are purely data-driven and lack the interpretability of physics-based models. In general, adding physical constraints could improve a model's prediction performance for extreme weather events (Hu & Li, 2024; Wang et al., 2024; Yan et al., 2024). Despite the lack of explicit physical constraints on the AI-generated TCs, it is an open question whether AI models can produce storm systems that are physically realistic. Therefore, we examine whether the Pangu and Aurora simulations satisfy the gradient wind balance (GWB), which is an important physical balance constraint for TCs. Specifically, above the boundary layer, mature TCs approximately satisfy the gradient wind balance between the pressure gradient force, centrifugal force, and Coriolis force in the azimuthal mean, as expressed in Equation (1).

$$g\frac{\partial Z}{\partial r} = \frac{V_\theta^2}{r} + fV_\theta \tag{1}$$

where $V_\theta$ is the azimuthal wind, $r$ is radius from the storm center, $f$ is the Coriolis parameter, $g$ is gravitational acceleration and $Z$ is geopotential height. By integrating the GWB equation from the storm center (i.e., $r = 0$) to a given radial distance $R$ (i.e., $r = R$). Eq. (1) can be reshaped as:

$$\int_0^R \frac{\partial Z}{\partial r}dr = [Z(R) - Z(0)] = \Delta Z = \frac{1}{g}\int_0^R (\frac{V_\theta^2}{r} + fV_\theta)dr \tag{2}$$

Here, we use the geopotential height deficit between the storm center and a given radial distance ($\Delta Z$) to evaluate whether the AI model satisfies GWB at each simulation timestep. Specifically, the simulated geopotential height deficit ($\Delta Z_{sim}$) is calculated by subtracting the azimuthally averaged $Z(R)$ from $Z(0)$. The geopotential field is one of the outputs of the Pangu/Aurora simulations and we can directly extract the azimuthal-average $Z(R)$ and $Z(0)$. Meanwhile, the theoretical geopotential height deficit under the gradient wind balance assumption ($\Delta Z_{GWB}$) can be computed based on Eq. (2) using the AI model wind field at 500 hPa pressure level. We use the wind field at the 500 hPa level to avoid deviations from GWB induced by the boundary layer, and to maintain consistency with previous gradient wind balance studies of AI weather models (Sun et al., 2024), enabling direct comparison with prior findings.

## 2.5 Extratropical Transition of TCs

We investigate if Pangu and Aurora can accurately simulate the process of extratropical transition (ET), which can be defined using the phase space criteria (Hart, 2003; Hart & Evans, 2001). ET starts when the TC transitions from a warm-core, nonfrontal (symmetric) cyclone to a warm-core, frontal (asymmetric) cyclone and ends when the cyclone transitions to a cold-core, frontal system that is a fully extratropical cyclone (ETC). The phase space is defined by two parameters: the cyclone thermal symmetry ($B$) and thermal wind ($-V_T^L$). The thermal symmetry parameter, $B$, distinguishes between symmetric and asymmetric (i.e., frontal) systems, and is calculated by averaging the 900–600 hPa layer thickness, measured within a 500 km radius to the right and left of the cyclone's direction of motion. The difference between these two averages indicates how thermally symmetric the system is, with positive values indicating warm advection downstream of the cyclone and an asymmetric structure, while negative values represents cold advection downstream of cyclone and a symmetric circulation (Hart, 2003; Hart & Evans, 2001). In practice, Hart (2003) suggests a threshold of 10m to distinguish symmetric vs asymmetric cyclones. The thermal wind parameter, $-V_T^L$, describes the change of geostrophic wind magnitude with height, and its sign indicates whether the system has a warm-core (positive) or cold-core (negative) in the vertical structure. First, the cyclone geopotential height difference ($\Delta Z = Z_{max} - Z_{min}$) within a 500-km radius is used to calculate the derivative of height difference with respect to pressure level (p), computed as $\partial(\Delta Z)/\partial \ln p$. Then, the parameter $-V_T^L$ is determined using a linear-regression fit of $\Delta Z$ between 900 and 600 hPa. Using $\boldsymbol{B}$ and $-\boldsymbol{V}_T^L$, storms can be classified into symmetric/warm-core, asymmetric/warm-core, symmetric/cold-core, or asymmetric/cold-core. The cyclone phase space has frequently been used as a diagnostic tool to identify the extratropical transition of TCs (Atallah & Bosart, 2003; Baker et al., 2022; Gori et al., 2023). Using the phase space criteria, we classify all historical storms as either tropical cyclone (TC; symmetric/warm-core), extratropical transition (ET; asymmetric/warm-core), or extratropical cyclone (ETC; asymmetric/cold core) based on their final phase over the storm lifetime. We then compare the relative frequency of TC, ET, and ETC events in ERA5 and Pangu/Aurora, and evaluate the spatiotemporal pattern of ET in ERA5, Pangu and Aurora.

## 3 Results and discussion

### 3.1 TC track densities

The TC track density maps for storms initialized when the IBTrACS wind speed first reaches 34 knots are generated for the NA and WNP basins, and the differences in AI-simulated track density and ERA5 track density are shown in Figure 2. Density difference maps between Pangu/Aurora and IBTrACS tracks are very similar to ERA5 (see Figure S1), since IBTrACS data is assimilated into ERA5. For the NA basin, there are 90 historical storm events and the maximum positive difference between Pangu and ERA5 is 9, while the maximum negative difference is -10. Aurora shows similar maximum difference (9 and -9). The Gulf of Mexico (GoM) has a high number of TCs (Fig. 2a,b), and the mean residual values over this region are -0.04 for Pangu and 0.01 for Aurora, indicating that both models have good agreement with ERA5 tracks. In the North Atlantic main development region (between 10º to 20º N and 20º to 70º W), the mean residual track density is 0.03 between Pangu and ERA5, and -0.36 between Aurora and ERA5. Generally, when considering the entire NA basin, the dark red and blue colors are randomly distributed. Therefore, there is no clear spatial trend in TC track density differences between the Pangu/Aurora model and ERA5. Rather, track differences across many storms appear to be due to random variations rather than systematic bias. The results are similar for the WNP basin (Fig. 2c,d), and the maximum positive and negative residuals are 12 and -19 for Pangu (Fig. 2c), and 15 and -13 for Aurora (Fig. 2d). A total of 97 storms are included in the WNP analysis, leading to higher cell values compared to the NA basin. The dark colors occur in regions where the total number of TCs is high (i.e., the Philippine Sea region of 120 ºE to 130 ºE). Similar to the NA basin, the mean residuals over the Philippine Sea are close to zero for both Pangu (0.06) and Aurora (-0.15). Moreover, the positive and negative residuals across the WNP basin appear to be evenly distributed. The FDR method (explained in section 2.3) is applied to both basins between Pangu and ERA5 and between Aurora and ERA5. The results return zero grid cells with significant differences under a false discovery rate equal to 5%. Therefore, although the Pangu and Aurora models may not, in general, generate track density maps that exactly match historical observations, across many storms the models generate storm track patterns that are realistic compared to the observed climatology.

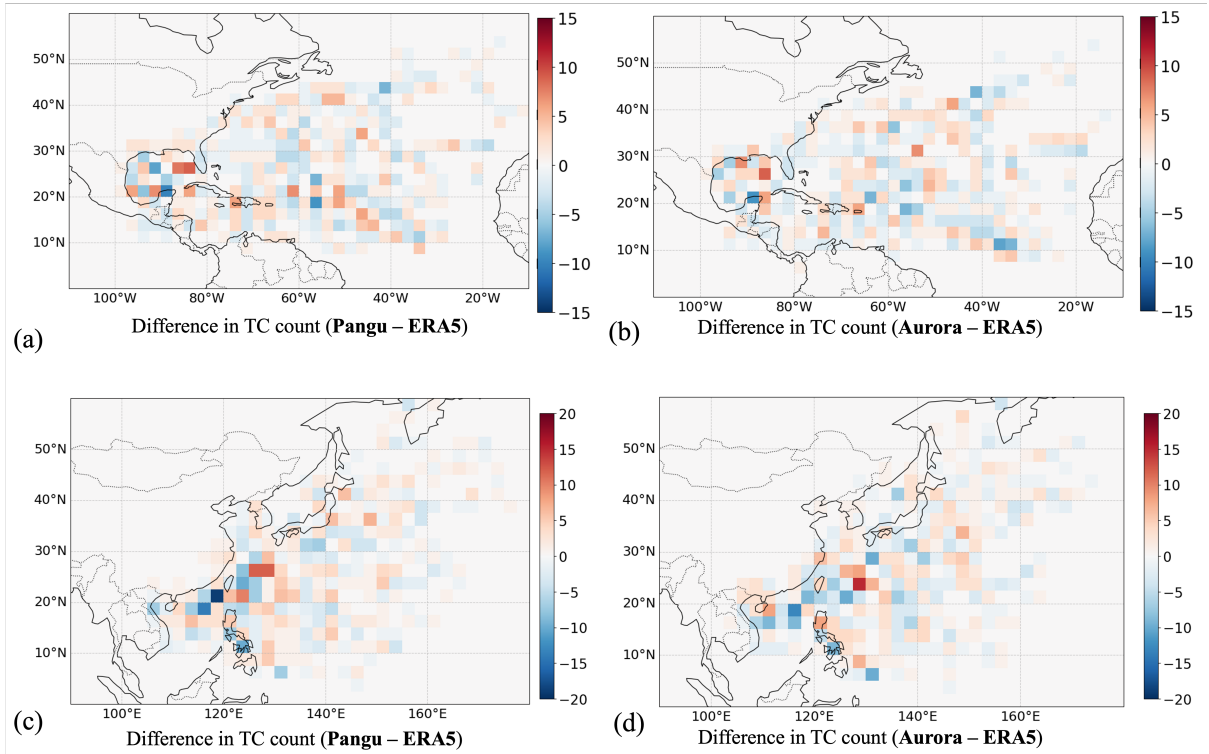

**Figure 2.** TC track density residual map at 2.5° latitude x 2.5° longitude resolution: (a) NA basin, Pangu -ERS5; (b) NA basin, Aurora -ERA5; (c) WNP basin, Pangu -ERA5; (d) WNP basin, Aurora -ERA5.

3.2 TC characteristics

After investigating the track densities produced by each AI model, we evaluate the distributions of various storm parameters and compare them to ERA5 and IBTrACS. Unless otherwise noted, all results presented here utilize TC simulations that were initialized when the storm first reached 34 knots, as recorded in IBTrACS.

Figure 3 shows the probability density functions of TC forward speed in the NA (Fig. 3a) and WNP (Fig. 3b) basins for two AI models and ERA5. In both basins, TCs most commonly travel at speeds of approximately 5-6 m/s. In the NA basin, the mean forward speeds are very close among ERA5 (5.84 m/s), Pangu (5.73 m/s), and Aurora (5.71 m/s). Similarly, in the WNP basin, the mean forward speeds are 5.56, 5.66, and 5.64 m/s for ERA5, Pangu, and Aurora, respectively. Notably, both models have similar densities as ERA5 for near 0 m/s forward speed, which indicates that Pangu and Aurora can realistically simulate the frequency of TCs that stall. Since the forward speed distributions of the AI models are very close to ERA5, we conducted a Kolmogorov–Smirnov (K–S) test to confirm statistical similarity (Supporting Information, Table S1) In the NA basin, both AI models and ERA5 appear to produce the same distribution of forward speed at the 5% significance level. In the WNP basin, there is a small but statistically detectable difference in forward speed between Aurora and ERA5. This is likely due to the long upper tail in Aurora's forward speed distribution in the WNP, and the heavy upper tail indicates that Aurora produces some TCs that move much faster (> 25m/s) than TCs in ERA5.

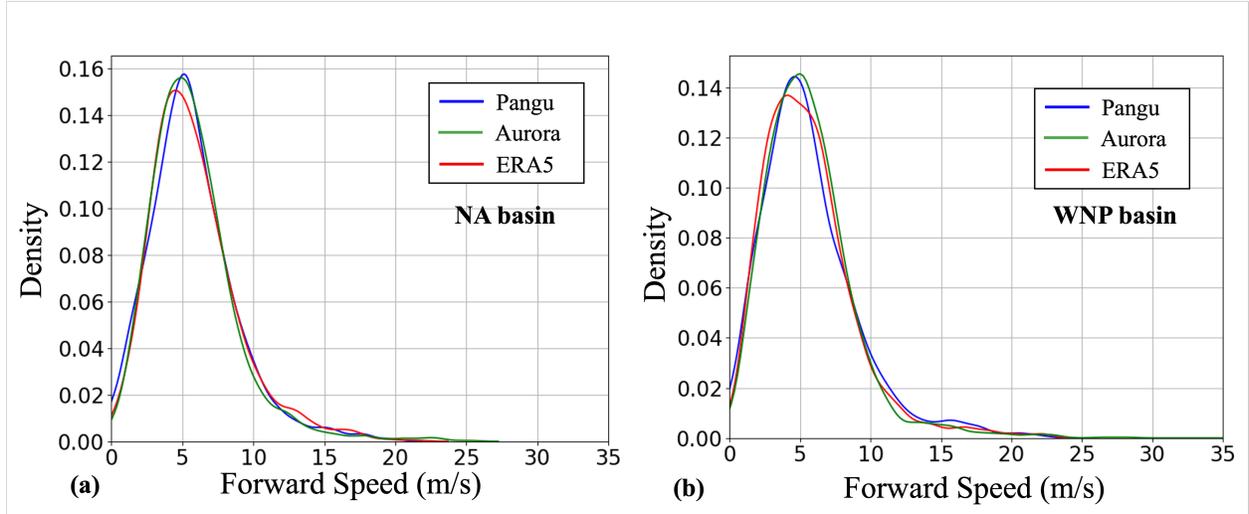

**Figure 3.** Probability density function of TC forward speed for the Pangu-Weather, Aurora Models and ERA5. The model simulation is initialized at the first timestep when $V_{max}$ reaches 34 knots. (a) NA basin. (b) WNP basin.

Besides forward speed, the remaining TC characteristics are analyzed using bivariate distributions, which facilitates evaluation of the dependence structure between different storm characteristics. We only include timesteps where the AI model or ERA5 predicted TC center is over the ocean, since the wind and pressure values over land would be impacted by the terrain and land surface characteristics.

Figure 4 shows the joint distribution of $V_{max}$ and MSLP across the two basins, with the main plot showing the joint distribution and rotated subplots showing the marginal distributions of $V_{max}$ (above the x-axis) and MSLP (to the right of the y-axis). In these marginal plots, Pangu, Aurora and ERA5 are overlaid for comparison. In the NA basin, Aurora and ERA5 produce very similar MSLP distributions, with median values of 1002.9 hPa and 1002.4 hPa, respectively, and similar shapes of the lower tail of the distributions. Pangu shows a slightly higher median (1004.6 hPa) than ERA5 and largely underestimates the lower tail. In the WNP basin, both AI models produce median MSLP values (994.8 hPa for Pangu and 992 hPa for Aurora) that closely match ERA5 (994.5 hPa), but while Aurora more realistically simulates the frequency of intense (low MSLP) TCs, Pangu underestimates their frequency. For example, ERA5 has 51 timesteps in the NA with MSLP less than 971 hPa and 50 timesteps in the WNP with MSLP less than 958 hPa. While Aurora simulates 30 timesteps below 971 hPa (in NA) and 37 timesteps below 958 hPa (in WNP), Pangu does not simulate any storms with MSLP less than these thresholds. Therefore, the Aurora model still slightly underestimates the frequency of the most intense storms, but it shows better performance than Pangu. K-S tests of the MSLP distributions confirm that Aurora and ERA5 are statistically similar in the NA basin but not the WNP basin, and Pangu is not statistically similar to ERA5 in either basin (Table S1). We implement a sensitivity test to investigate whether the initial vortex intensity used to initialize the Pangu model changes the subsequent distribution of MSLP. We conduct additional Pangu simulations for all TCs initialized when/if each storm first reaches 25 knots, 50 knots, and 64 knots (as recorded in IBTrACS). Despite starting the Pangu simulations with higher initial TC intensity, we observe that the model still does not capture the lower tail of the MSLP distribution (Figures S2-S4). Conversely, initializing the model at a lower initial intensity does not produce significantly more biased MSLP estimates (Figure S2).

Similar performance is observed for the $V_{max}$ distribution, where Pangu underpredicts high wind speeds (>25 m/s) but Aurora captures them well compared to ERA5. In the NA basin, Aurora and ERA5 have similar median $V_{max}$ values (12.4 m/s and 12.6 m/s, respectively), whereas Pangu shows a lower median of 10.8 m/s. In the WNP basin, the median $V_{max}$ values are 13.7, 15.5, and 14.4 m/s for Pangu, Aurora, and ERA5, respectively. As with the MSLP distributions, K-S tests (Table S1) confirm that Aurora and ERA5 are statistically similar in the NA basin but not the WNP basin and Pangu is not similar to ERA5 in either basin. Comparing Figure 4(a) and (c) or Figure 4(d) and (f), it is clear that Pangu has higher density than ERA5 at high MSLP and low $V_{max}$ regions and lower density than ERA5 at low MSLP and high $V_{max}$, indicating an overall bias in the intensity of Pangu-simulated TCs. However, Aurora can simulate the shape of the entire joint distribution more accurately than Pangu. The shape of the joint distribution found here agrees with past studies investigating TC wind-pressure relationships (Chavas et al., 2017; Knaff & Zehr, 2007), which also found an inverse quadratic relationship between MSLP and $V_{max}$. Overall, the similar coefficients of the best fit quadratic regression lines between Pangu/Aurora and ERA5 in each basin indicate that the AI models can accurately simulate the MSLP-$V_{max}$ covariance relationship even if they do not reproduce the full intensity spectrum. These findings regarding TC intensity in AI models align with previous research by Sun et al. (2024), which showed that AI weather models can be unreliable for Category 3-5 TCs. It is important to note that TC intensity in ERA5 is also underpredicted compared to observations from IBTrACS due to ERA5's 25 km horizontal resolution, which cannot fully resolve intense storms (Bian et al., 2021; Gori et al., 2023). Therefore, the Pangu TCs would be even more biased when compared against observations. The comparison between ERA5 reanalysis and IBTrACS for $V_{max}$-MSLP distribution is presented in Figure S5, which shows that ERA5 underestimates the intense TCs and cannot resolve wind speeds greater than 30 m/s. Therefore, even though Aurora has improved intensity simulation, it is still highly biased compared to observations.

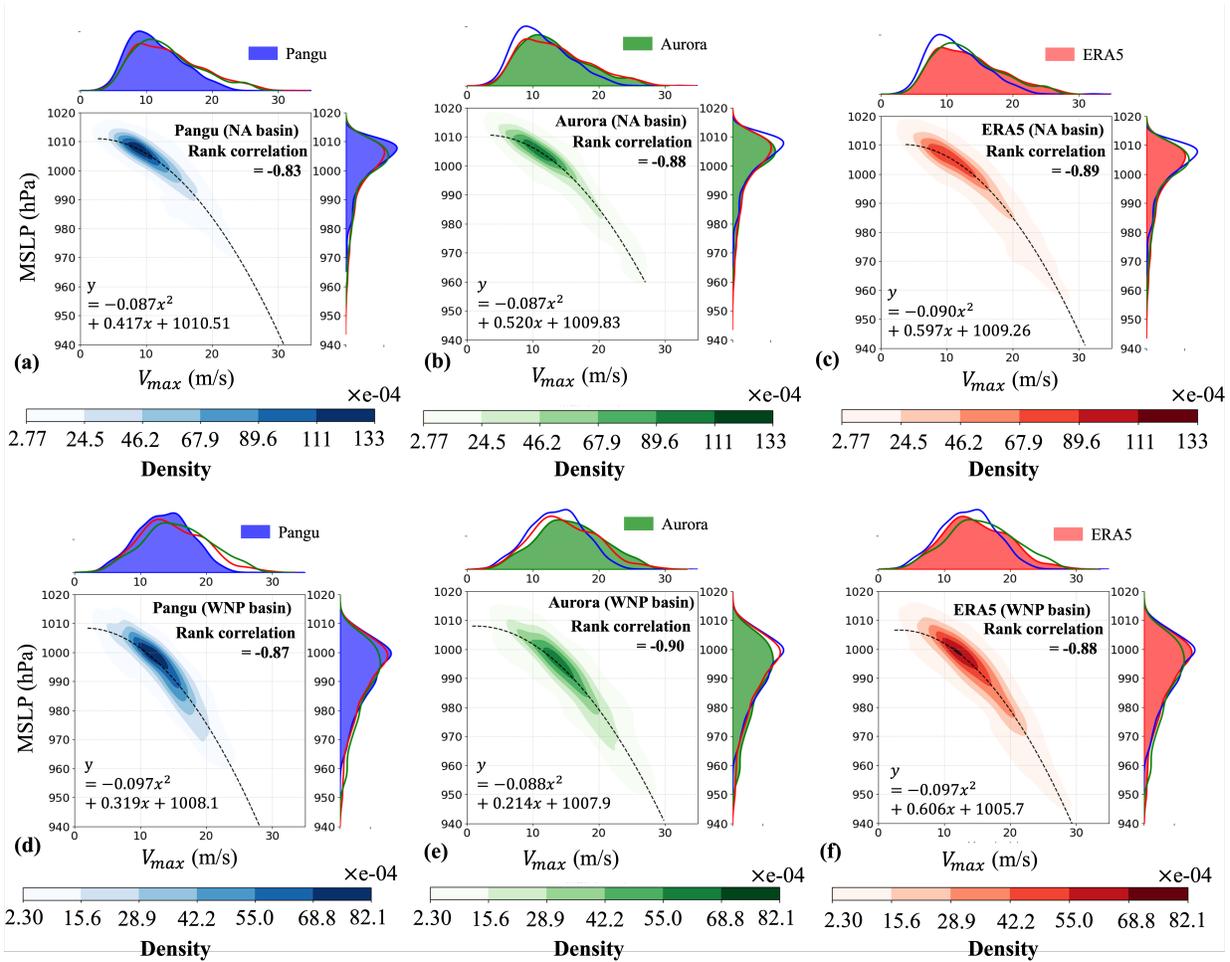

**Figure 4.** Joint distribution of $V_{max}$ and MSLP across Pangu-Weather, Aurora Model and ERA5 in the (a-c) NA and (d-f) WNP basin. Data is filtered as described in the text.

To analyze the size climatology of Pangu/Aurora vs ERA5 TCs, the joint distributions of $R_{max}$ and outer size $r_8$ are shown in Figure 5. $R_{max}$ refers to the radius of peak azimuthal-mean azimuthal wind speed while $r_8$ refers to the outer radius at which the azimuthal-mean azimuthal wind first drops below 8 m/s. The same TC time points are presented in Figures 4 and 5, with TCs over land being excluded from the analysis. In the NA basin, the distributions for the $r_8$ shows good agreement, with median $r_8$ values similar between Pangu (367 km), Aurora (367 km) and ERA5 (345 km). The notable difference between models appears in the WNP $r_8$ distribution, where ERA5 has a smaller median $r_8$ of 397 km compared to Pangu, which is close to 450 km, and Aurora, which is 435 km. Overall, the density function and K-S test results (Table S1) indicate Aurora/Pangu and ERA5 have good agreement for $r_8$, with Aurora demonstrating even better agreement than Pangu. However, in contrast to $r_8$, the Pangu and Aurora distributions of $R_{max}$ shows consistent biases in both basins, although the bias for Aurora is smaller. The median ERA5 $R_{max}$ is 112.5 km in the NA basin and 105 km in the WNP. However, Pangu consistently produces larger inner sizes in both basins (median $R_{max}$ of 150 and 143 km in NA and WNP basin). The Aurora model has a median $R_{max}$ of 130 km and 117 km for NA and WNP basin, respectively. The $R_{max}$ distributions of Aurora/Pangu are not statistically similar to ERA5 in either basin. Comparing the joint distributions in Figure 5(a), (b) and (c), the shapes of the $R_{max}$ - $r_8$ distributions are similar between Pangu/Aurora and ERA5, but the Pangu distributions are noticeably shifted upward (higher $R_{max}$). This indicates that while Pangu accurately simulates the climatology of the TC outer size, it overestimates the corresponding $R_{max}$ while Aurora has better estimation for the $R_{max}$-$r_8$ joint distribution. This pattern persists in the WNP basin based on Figure 5(d), (e) and (f). The results are similar when using the outer size metric $r_6$, rather than $r_8$ (Figure S6), which also indicates outer size $r_8$ and $r_6$ are highly correlated and the AI weather models can accurately simulate the outer TC wind field ($r_8$ and $r_6$).

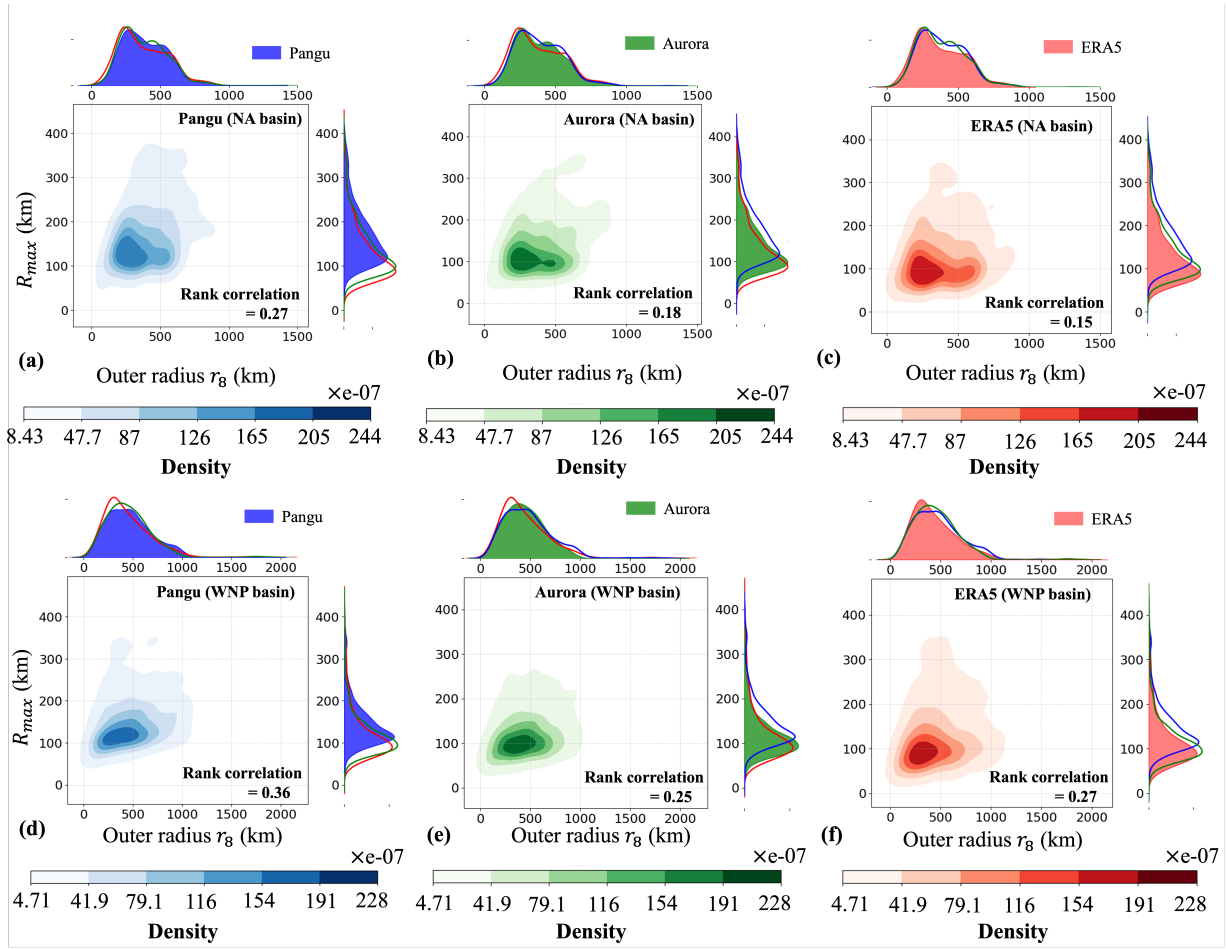

**Figure 5.** Joint distribution of TC outer size and $R_{max}$ across Pangu-Weather, Aurora Model and ERA5 in the (a-c) NA and (d-f) WNP basin. Data is filtered as described in the text.

The dependence structure between $V_{max}$ and $R_{max}$ is discussed using joint distributions shown in Figure 6. Although there can be large variability in $R_{max}$ across storms of similar intensity, small $R_{max}$ is generally associated with higher intensity and/or intensifying storms, while large $R_{max}$ is generally associated with weak and/or weakening storms (J. Xu & Wang, 2015) Therefore, we would expect to observe an inverse relationship between $V_{max}$ and $R_{max}$ in both the ERA5 and Pangu/Aurora TCs. However, comparing Figure 6(a) and (c), we find that while there is a definite inverse exponential relationship between $V_{max}$ and $R_{max}$ in ERA5 (which is more pronounced in the NA basin than WNP basin), the Pangu model does not fully reproduce this relationship, especially in the WNP basin. Specifically, the Pangu model results are highly clustered in the region where $V_{max}$ is around 10 m/s and $R_{max}$ is around 100 to 200 km, and simulates very few storms with $R_{max}$ less than 100 km, which is unrealistic. For ERA5 data, the high-density region is more spread out where $V_{max}$ is from 10 to 15 m/s and $R_{max}$ ranges from 80 km to 130 km. For the Aurora simulation, the inverse exponential relationship is clearer in the NA basin (less so in the WNP basin) and the simulation results cluster in the region very close to ERA5 data. The WNP results in Figure 6(d) and (f) show the same trend, where $V_{max}$ is more spread out and reaches higher wind speeds in Aurora. As observed in Figure 6, the Pangu model generally simulates the peak wind speed at larger radius and lower magnitude than ERA5, producing TCs that are too weak and that have inner cores that are too large compared to reanalysis. As for Aurora TCs, the model can produce a more reasonable representation of the inner region of the storm with a smaller bias for intense storms, but still tends to overestimate $R_{max}$. Figure S7 presents the joint distribution of $V_{max}$ and $R_{max}$ from IBTrACS compared with ERA5 data. The inverse exponential relationship between $V_{max}$ and $R_{max}$ using IBTrACS is clear and consistent across both basins, indicating that both ERA5 and the AI models struggle to reproduce an accurate $V_{max}$-$R_{max}$ dependence structure. The Pangu model, in particular, does not resolve the inverse $V_{max}$-$R_{max}$ relationship well, especially in the WNP basin.

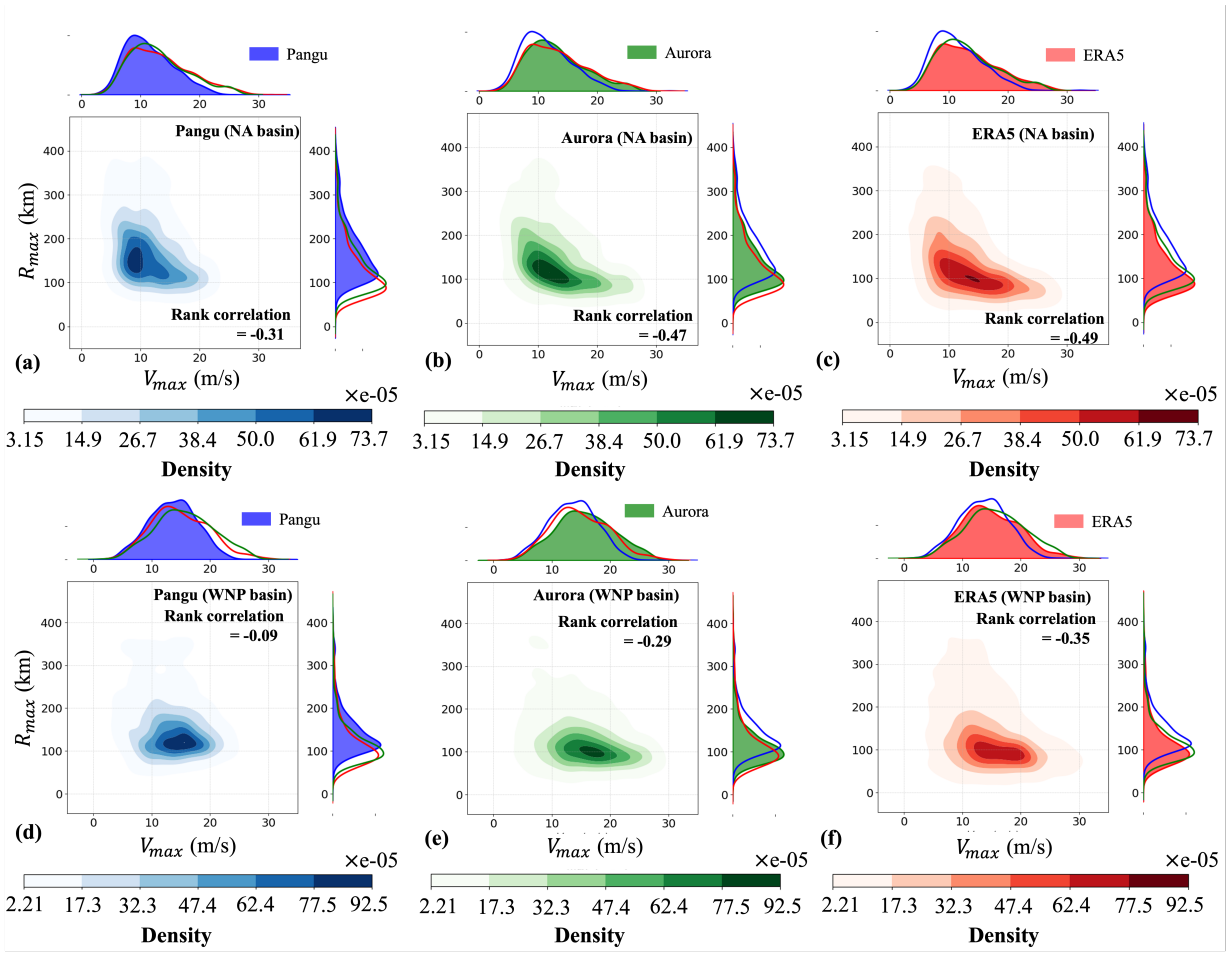

**Figure 6.** Joint distribution of $V_{max}$ and $R_{max}$ across Pangu-Weather, Aurora Model and ERA5 in the (a-c) NA and (d-f) WNP basin. Data is filtered as described in the text.

In general, Pangu and Aurora realistically represent most aspects of the dependence structure between storm characteristics, with the exception of the $R_{max}$-$V_{max}$ relationship. For example, in the NA basin both AI models and ERA5 show strong negative rank correlations between MSLP and $V_{max}$ (-0.83 for Pangu, -0.88 for Aurora, and -0.89 for ERA5), which aligns with the expected physical relationship where lower central pressures correspond to higher wind speeds. However, while ERA5 indicates a moderate negative rank correlation between $R_{max}$ and $V_{max}$ (-0.49), and Aurora reproduces this correlation (-0.47), Pangu shows weaker relationship (-0.31). Similarly, ERA5 has a moderate correlation of 0.42 between $R_{max}$ and MSLP (see Figure S8), which is slightly weaker in Aurora (0.37) and much weaker in Pangu (0.21). For the WNP basin, similar discrepancies exist in the $R_{max}$-$V_{max}$ relationship, with Pangu showing very weak correlation (-0.09) compared to ERA5's correlation of -0.35 and Aurora's correlation of -0.29. WNP TCs in ERA5 have a weaker correlation between $R_{max}$ and MSLP (0.20), and even weaker correlations in Aurora (0.15) and Pangu (-0.08). The rank correlations between storm variables (Fig. S8), combined with Figures 4 and 6, indicate that Aurora slightly underestimates the tendency for more intense storms to have smaller radius of maximum winds, while Pangu more significantly underestimates this tendency. Overall, these results indicate that while Pangu and Aurora both capture large-scale relationships between storm intensity and outer size, capturing the variability and co-variability of inner size ($R_{max}$) is more challenging. This limitation should be considered when using AI-generated TCs for hazard assessments, as it may lead to unrealistic combinations of storm parameters that could affect rainfall and storm surge predictions

Finally, we investigate whether there are any temporal trends in the model biases for $V_{max}$, $R_{max}$, and MSLP. Previous studies have found that AI models tend to develop a low intensity bias during the first day of TC simulations (DeMaria et al., 2025). Therefore, we subset our analysis by the number of days since the start of the simulation (i.e., 1-day, 2-day, 3-day). For each subset, the mean relative bias between Pangu/Aurora and ERA5 is calculated for MSLP, $R_{max}$, $r_8$, and $V_{max}$ (Figures S9-S12). As shown in Figure S9, the relative bias for MSLP generally increases with simulation day for Pangu and Aurora in the NA basin, indicating a weakening of TCs over time compared to ERA5. In the WNP, Aurora has an increasingly negative mean bias over time, indicating a relative strengthening of TCs compared to ERA5. Similarly, Figure S11 indicates that while both Pangu and Aurora are biased in estimating $V_{max}$, Pangu underestimates $V_{max}$ on average, while Aurora overestimates it. Most importantly, when the data are split into two groups (1–4 days and 5–7 days), a clear trend can be seen that predictions are more accurate in the first group than in the second. This is likely because both AI models use autoregressive inference, in which the output of one timestep is used as input for the next, causing errors to accumulate as lead time increases. In contrast to temporal trends in intensity bias, neither AI model displays a clear trend in the outer or inner size bias over time (Figures S10-S12).

### 3.3 Gradient Wind Balance of TCs

The variables $\Delta Z_{sim}$ and $\Delta Z_{GWB}$, as defined in Section 2.4, are used to evaluate whether the GWB is satisfied. Data points are first filtered using three criteria: 1) simulations are initialized when $V_{max}$ first reaches 34 knots in IBTrACS, 2) only timesteps where the AI-predicted TC center is over the ocean are retained, and 3) timesteps with $V_{max}$ smaller than 8 m/s in the AI model are excluded to avoid weak, disorganized storms. Figures 7 present hexbin plots (color corresponds to density of points within the hexagonal bin) of $\Delta Z_{sim}$ vs. $\Delta Z_{GWB}$ for both basins. Based on Equation (2), $\Delta Z_{GWB}$ is computed by integrating from the storm center ($r = 0$) to an upper radial limit ($r = R$). Two values of $R$ are selected to assess the GWB constraint in both the inner and outer regions of the 500 hPa azimuthal wind profile: 100 km (Figure 7a-c, g-i) for the inner region and 500 km (Figure 7d-f, j-l) for the outer region. In the NA basin, ERA5 generally conforms well to GWB in both the inner (R²=0.80) and outer (R²=0.93) regions. In contrast, Pangu exhibits a weak correlation between the simulated ($\Delta Z_{sim}$) and GWB ($\Delta Z_{GWB}$) height difference at $R$ equals 100 km ($R^2 = 0.23$), while Aurora shows slightly improved performance ($R^2 = 0.44$). The median line for both Pangu and Aurora deviates substantially from the 1:1 line compared to ERA5. Specifically, Pangu shows a relative mean bias (i.e., mean of $\Delta Z_{sim}$ - $\Delta Z_{GWB}$ over mean $\Delta Z_{GWB}$) of -0.4, while Aurora shows a negative mean bias of -0.38, and visually deviates largely from GWB for intense storms. As the radial distance increases to 500 km, correlations between $\Delta Z_{sim}$ and $\Delta Z_{GWB}$ improve substantially for both AI models ($R^2 = 0.90$ for Pangu and Aurora), closely matching ERA5 ($R^2 = 0.93$). The median line of Pangu and Aurora follows the 1:1 line more closely at R = 500km, although Aurora still diverges from GWB for high intensity storms. These results suggest that AI models fail to satisfy the GWB constraint in the inner region of the TC but perform better in the outer region. ERA5, on the other hand, generally satisfies the GWB constraint across both regions.

A similar pattern is observed in the WNP basin (Figure 7 g-l). At 100 km, Aurora shows slightly worse performance with a higher relative mean bias (-0.29), compared to Pangu (-0.26).

At 500 km, both AI models and ERA5 demonstrate strong agreement with GWB ($R^2 = 0.93$ for Pangu, $R^2 = 0.94$ for Aurora and $R^2 = 0.95$ for ERA5), and the relative mean bias drops to -0.08 for Pangu and -0.11 for Aurora. Additionally, ERA5 maintains a lower relative mean bias in both the inner (-0.23 for 100 km) and outer (0.10 for 500 km) regions. For both basins, the median line in AI models' 100 km results deviates significantly from the 1:1 line at high $\Delta Z_{sim}$ (which corresponds to low MSLP), indicating a potential systematic deviation from the GWB-based pressure deficit in intense storms' inner cores. The wider 5%-95% ranges in Pangu (Figures 7a, g) and Aurora (Figures 7b, h) imply greater uncertainty and variability in the simulated height difference for the inner core of storms, suggesting that AI-based wind field predictions may not be reliable for the inner storm region. The AI models' improved performance at 500 km suggests they capture the large-scale environment better than the inner-core dynamics. Our results align well with the earlier study (Sun et al., 2024), confirming that AI models struggle to satisfy gradient wind balance in TC inner cores.

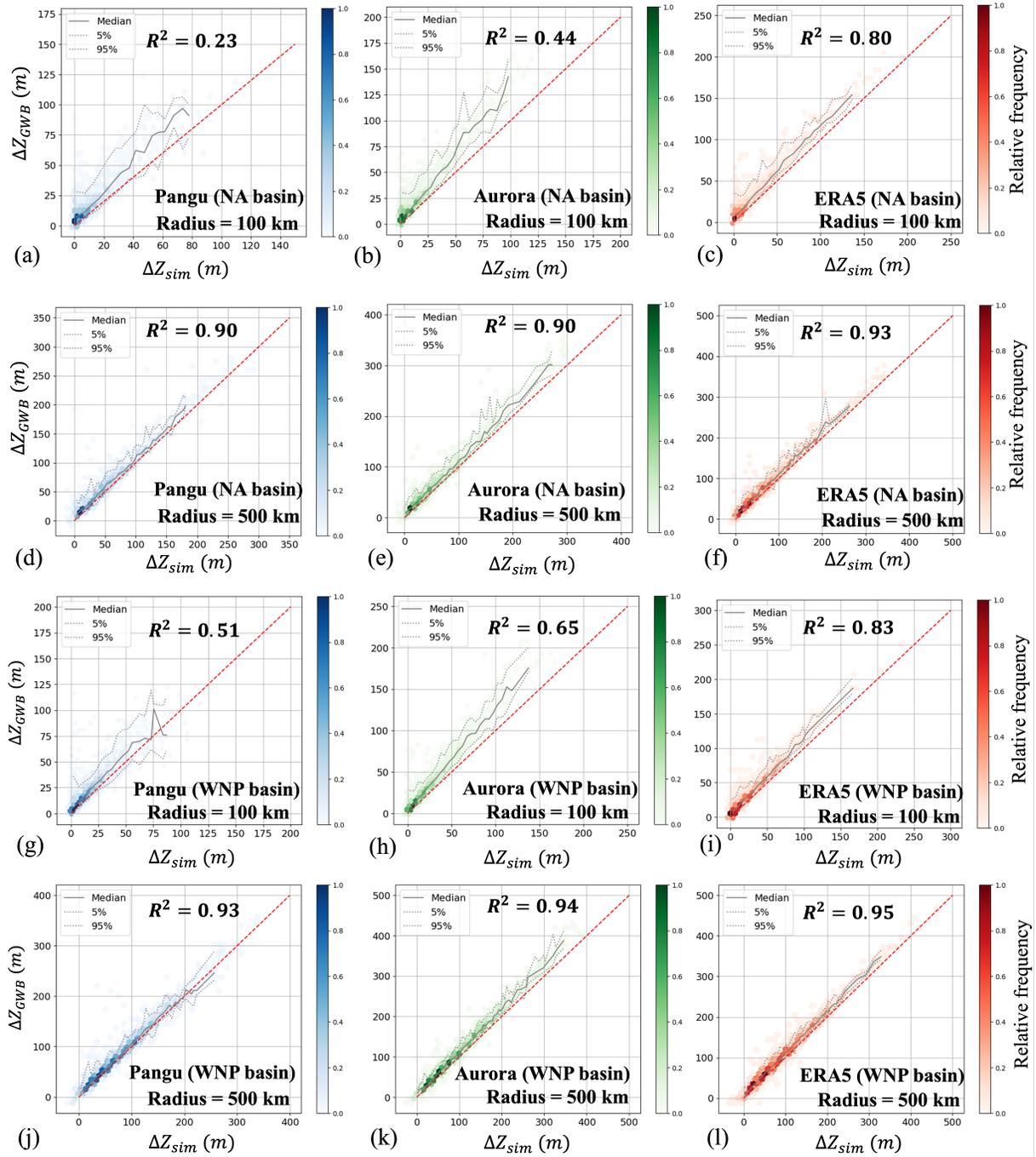

**Figure 7.** Theoretical vs. simulated geopotential height deficit $\Delta Z$ across Pangu, Aurora and ERA5. (a-c) Results at 100 km radial distances in NA basin. (d-f) Results at 500 km radial distances in NA basin. (g-i) Results at 100 km radial distances in WNP basin. (j-l) Results at 500 km radial distances in WNP basin.

3.4 Extratropical Transition

To investigate the ERA5 and AI model climatology of extra-tropical transition, the cyclone phase space parameters described in Section 2.4 are calculated for each timestep of each TC simulation, and the TC is classified as tropical cyclone (TC), extratropical transition (ET), extratropical cyclone (ETC), or unclassified. If all the time steps for a given TC have nonfrontal warm core (i.e., $B$ smaller than 10 and $-V_T^L$ larger than 0), the simulation is classified as a TC. If the simulation includes at least one transition from nonfrontal warm core to frontal warm core (i.e., $B$ larger than 10 and $-V_T^L$ larger than 0) and maintains this state for more than two timesteps (12 hours), but does not yet develop into a frontal cold-core system (i.e., B larger than 10 and $-V_T^L$ less than 0), the storm is classified as ET. Finally, if the simulation starts from nonfrontal warm core and eventually transitions to frontal cold core, it is classified as ETC. Any cases that do not satisfy these conditions are unclassified. Unclassified cases may be timesteps that fall into the symmetric cold core space at the end of the simulations or initialize as a symmetric cold core and directly transform to an asymmetric warm core or cold core. Unclassified cases could also be due to incorrect calculation of phase parameters due to inaccurate simulation of the storm's direction of motion or geopotential height between 900 and 600 hPa. However, most storms can be successfully classified into the TC, ET, and ETC categories.

Table 1 shows the number of storms in ERA5, Aurora and Pangu that fall into each category for the NA and WNP basins. Table 1 shows that across many storms, there is good agreement between AI models and ERA5 regarding the relative frequency of TC, ET, and ETC occurrence. In the NA basin, ERA5 produces the largest number of ET/ETC storms (36 cases total), while Pangu and Aurora have fewer ET/ETC events (24 for Pangu and 29 for Aurora). The number of unclassified storms is lowest in ERA5 (8 cases), compared to 11 cases in both Pangu and Aurora, indicating that both AI models occasionally produce phase trajectories that may be inconsistent with physically-realistic transitions. In the WNP basin, ERA5 has 48 cases classified as ET/ETC while Aurora (45 cases total) and Pangu (38 cases total) slightly underestimate the total number of ET/ETC. Similar to NA basin, ERA5 has the lowest unclassified cases in the WNP basin. In both basins, for Category 3-5 major hurricanes, such as Eta (2020) in NA basin and Surigae (2021) in WNP basin, ERA5 classifies them as TC but the Pangu model classifies them as ET/ETC. As discussed in the previous section, the Pangu model cannot reproduce these intense TCs. Therefore, it is likely that the phase parameters are unrealistic compared to ERA5. In contrast, the Aurora model accurately classifies most of these cases because it has better performance in simulating intense storms.

| Model | Tropical Cyclone (TC) | Extratropical Transition (ET) | Extratropical Cyclone (ETC) | Unclassified |
|---|---|---|---|---|
| North Atlantic Basin | | | | |
| Pangu | 55 | 11 | 13 | 11 |
| Aurora | 50 | 18 | 11 | 11 |
| ERA5 | 46 | 18 | 18 | 8 |
| Western Pacific Basin | | | | |
| WNP: Pangu | 50 | 17 | 21 | 9 |
| WNP: Aurora | 45 | 21 | 24 | 7 |
| WNP: ERA5 | 45 | 21 | 27 | 4 |

**Table 1.** Counts of Tropical Cyclone, Extratropical Transition, and Extratropical Cyclone Cases in Pangu, Aurora, and ERA5.

      Figure 8 illustrates the phase space evolutions of one example storm, Hurricane Hagupit (2020), which transitioned from TC to ETC. Figure 8 (a) plots Hurricane Hagupit's track, while Figures 8(b-d) show the phase parameters for each timestep in ERA5, Aurora, and Pangu. The dashed red lines indicate the boundary of each phase. The ERA5 reanalysis of Hagupit's track starts in the symmetric warm core phase and first reaches the asymmetric phase at point A (onset of ET). After point A, the storm starts the transition to an extratropical cyclone. The storm becomes fully extratropical after point B. Both Aurora and Pangu simulate the onset of ET (points C and E, respectively) at a similar location as ERA5 (near 35° N). However, the transition to a fully extratropical cyclone occurs at different locations, marked by rectangles. The AI models and ERA5 shows similar latitudes of transition completion (39.25°N for Aurora, 41.5°N for ERA5, and 42.5°N for Pangu). However, they show large difference in the longitude. These differences likely reflect model limitations near the end of the 7-day simulation window, when accumulated prediction errors become more significant. Based on Table 1 and Figure 8, we conclude that AI models can realistically simulate the frequency of ET, and the location of ET onset, but may not simulate the full transition from ET to ETC accurately.

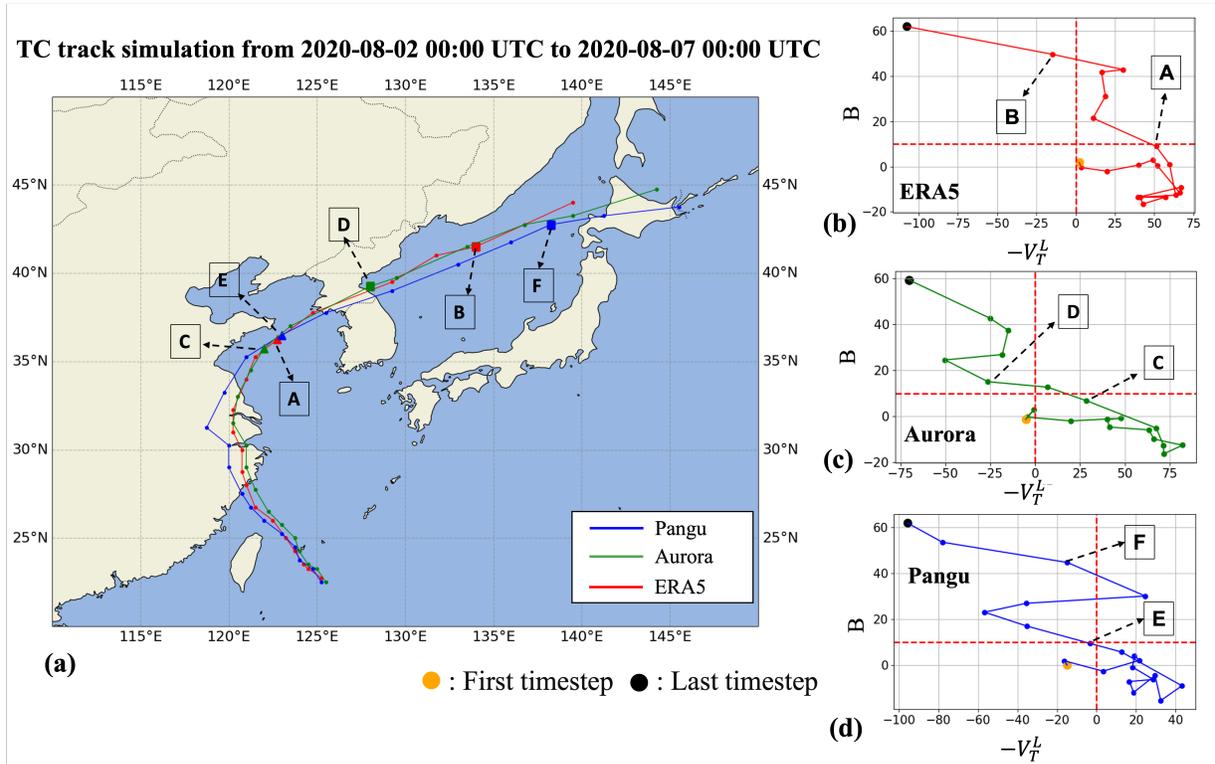

**Figure 8.** Illustration of extratropical transition in Hurricane Hagupit (2020). (a). TC track plot. (b, c, d). Phase parameters for Pangu, Aurora and ERA5, respectively.

## 4 Conclusions

This study presents a comprehensive climatological benchmark of AI-generated TCs using the Pangu-Weather and Aurora models. Through systematic comparisons with ERA5 reanalysis of TC events in the North Atlantic and Western North Pacific basins from 2020 to 2025, we assessed key TC characteristics including track density, forward speed, minimum pressure, storm size, and extratropical transition. The results reveal that the Pangu model can reasonably capture the climatology of TC outer size and forward speed, as well as the climatology and spatiotemporal evolution of extratropical transition. However, it cannot reproduce the climatology of minimum central pressures and maximum sustained wind speeds of TCs and is biased for intense events. Also, it systematically overestimates the radius of maximum wind, leading to storms that are too large and too weak. Aurora outperforms Pangu in reproducing TC climatology, particularly for intense storms. It reproduces the climatology of minimum central pressure and maximum sustained wind speed with only minor biases for the most intense events. Aurora also has improved simulation of the radius of maximum wind, with median $R_{max}$ values closer to ERA5, but still overpredicted. Moreover, it maintains high accuracy in parameters where Pangu already performs well, including track density, forward speed, outer size, and the spatiotemporal evolution of extratropical transition. There are several reasons that could explain Aurora's improved performance over Pangu in TC simulations. First, the two models adopt distinct architectures: Pangu relies on a 3D Earth-specific transformer tailored to ERA5 atmospheric fields, whereas Aurora combines Perceiver-based encoders/decoders with a 3D Swin Transformer processor that allows greater flexibility in representing multi-scale processes. Second, the Aurora model architecture is substantially larger in scale (around 1.3 billion parameters versus around 64 million in Pangu), providing greater capacity to capture the nonlinear dynamics of tropical cyclones. Third, the training strategies diverge between the two models. Specifically, Pangu was trained only on ERA5 reanalysis, while Aurora was pretrained on over one million hours of heterogeneous Earth system data, including ERA5, CMIP6, IFS forecasts, and GFS data. Subsequently, Aurora was fine-tuned with IFS-HRES data, which offers finer spatial resolution and is particularly effective for tropical cyclone track and intensity prediction. These architectural advances, larger capacity, and richer multi-source training pipeline might explain Aurora's superior performance relative to Pangu. Despite improved intensity simulation in Aurora, the advances in model architecture and training data still do not guarantee that simulated TCs are physically realistic. Specifically, both AI models show large uncertainty and variability in the simulated geopotential height difference for the inner core of storms and fail to simulate geopotential height differences that satisfy the gradient wind balance theory. However, at the outer region of the storm, the AI models' performance increases and can satisfy GWB. Given that the 34-knot threshold yields acceptable TC simulations and represents an early enough stage in TC development, it can serve as a practical initialization timestep for starting the TC simulations.

The climatological benchmarking presented here can serve as a guide for future studies and risk assessment efforts using AI-based weather models. By identifying systematic biases (e.g., the underestimation of TC minimum pressure), this benchmarking study enables targeted bias correction strategies that can improve the reliability of downstream applications. One promising direction is to generate synthetic TC events using AI models (Jing et al., 2024), then apply corrections based on the biases quantified here to create more realistic storm wind fields. Both AI models are biased for the most intense storms (both MSLP and $V_{max}$) while failing to satisfy physical constraints. Future work could utilize quantile-mapping approaches (Cannon et al., 2015) to bias correct the intensity of simulated Pangu/Aurora TCs. Since the $V_{max}$, MSLP and size of TCs are physically related (Chavas et al., 2017), future bias correction approaches should correct the Pangu/Aurora inner size conditioned on the corrected $V_{max}$ to ensure physically realistic storms. Here, we find that the outer TC size from AI models is unbiased compared to ERA5 but the inner size ($R_{max}$) from Pangu is biased high (too large). Past studies indicate the inner-core circulation and outer circulation are controlled by different physics (Chavas et al., 2013; Irish et al., 2008; Lin & Chavas, 2012), suggesting that the inner size could be bias-corrected independently from the AI-simulated outer size. To apply these AI models in future TC research in a trustworthy way, the biases inherent to each storm parameter (and their correlations) need to be corrected to satisfy physical assumptions. Therefore, future work could correct the AI wind field parameters using existing physics-based wind models (e.g., (Chavas et al., 2015)) to develop a corrected complete wind profile that would conform to gradient wind balance. Future work should also examine whether AI simulated TCs (and associated bias corrections) can also support a range of follow-up research, including enabling computationally-efficient simulation of high-resolution rainfall prediction, which relies on accurate wind field input, and subsequent coastal impact assessments. Ultimately, integrating bias-corrected AI-generated TCs into risk frameworks could enhance our ability to evaluate future hazard scenarios under climate change and improve disaster preparedness strategies.

## Acknowledgments

We thank the three anonymous reviewers for helping improve the manuscript. This study is supported by National Science Foundation (NSF) CHIRRP project award number 2440167.

## Conflict of Interest

The authors declare there are no conflicts of interest for this manuscript.

## Data Availability Statement

The Pangu-Weather model and pretrained weight can be downloaded from the Github (https://github.com/198808xc/Pangu-Weather). The Aurora model and pretrained weight can also be downloaded from Github (https://github.com/microsoft/aurora). The tropical cyclone best track data used in this study can be downloaded from the National Hurricane Center website (https://ncics.org/ibtracs/). The ERA5 reanalysis data is available at the European Centre for Medium-Range Weather Forecasts (https://www.ecmwf.int/en/forecasts/dataset/ecmwf-reanalysis-v5).